\documentclass[11pt,a4paper]{article}
\usepackage{graphicx}
\usepackage{amsmath}
\usepackage{cite}
\usepackage{amsfonts}
\usepackage{amssymb}
\newtheorem{theorem}{Theorem}

\newtheorem{lemma}{Lemma}

\newenvironment{proof}[1][Proof]{\textbf{#1.} }{\ \rule{0.5em}{0.5em}}
\begin{document}

\title{Total and Partial Amplitude Death in Networks of Diffusively Coupled Oscillators}
\author{Fatihcan M. Atay\medskip\\Max-Planck-Institute for Mathematics in the Sciences\\
Inselstr. 22, Leipzig 04103, Germany\\
e-mail: \tt{atay@member.ams.org}
}
\date{}
\maketitle
\begin{abstract}
Networks of weakly nonlinear oscillators are considered with diffusive and
time-delayed coupling. Averaging theory is used to determine parameter ranges
for which the network experiences amplitude death, whereby oscillations are
quenched and the equilibrium solution has a large domain of attraction. The
amplitude death is shown to be a common phenomenon, which can be observed
regardless of the precise nature of the nonlinearities and under very general
coupling conditions. In addition, when the network consists of dissimilar
oscillators, there exist parameter values for which only parts of the network
are suppressed. Sufficient conditions are derived for total and partial
amplitude death in arbitrary network topologies with general nonlinearities,
coupling coefficients, and connection delays.
\end{abstract}



\section{Introduction}

The interaction of oscillatory systems arise naturally in many areas of
science and has intrigued investigators for a long time. The earlier studies
go back to the 17th century when the Dutch scientist Christiaan Huygens
observed that two clocks mounted on a common basis tend to synchronize
regardless of the initial conditions \cite{Huygens,Bennett02}. In the last
several decades, models based on coupled oscillators have found diverse
applications in physics and biology. Among many examples are arrays of
Josephson junctions \cite{Hadley88a,Hadley88b}, semiconductor lasers
\cite{Varangis97,Hohl97}, relativistic magnetrons \cite{Benford89}, chemical
reactions \cite{Schreiber82,Kuramoto84b,Crowley89,Dolnik96}, circadian
pacemakers \cite{Daan78,Kawato80}, intestinal electrical rhythms
\cite{Linkens77}, and a variety of biological processes
\cite{Winfree80,Holstein-Rathlou87,Murray93}. The investigation of these
systems has provided striking examples of different types of dynamical
behavior that can be induced by the presence of coupling. Synchronization is
probably the most widely studied among these
\cite{Shinomoto86,Bonilla87,Pikovsky-book01}. Another interesting dynamics is
the so-called amplitude (or oscillator) death, whereby individual oscillators
cease to oscillate when coupled and go to an equilibrium solution instead. One
of the first accounts of this phenomenon was given in the 19th century by Lord
Rayleigh, who noted that when two organ pipes stand side by side they
``...~may almost reduce one another to silence'' \cite[Vol.~2, Chapter
XVI]{Rayleigh45}. Later, amplitude death was observed in electronic systems
\cite{Copeland66,Copeland67} and chemical oscillators \cite{Bar-Eli85},
causing renewed interest in this area. A theoretical analysis was given in
\cite{Aronson90}, which considered a pair of linearly coupled oscillators each
of which is near a Hopf bifurcation. In particular, it was established that a
fairly general class of such systems can exhibit amplitude death in case the
individual oscillator frequencies are sufficiently different and the coupling
is diffusive. In a similar vein, it was shown that a network of limit cycle
oscillators can experience amplitude death if there is a distribution in
frequencies \cite{Shiino89,Ermentrout90,Mirollo90}.

A series of more recent studies have demonstrated that time delays in the
coupling can induce amplitude death even in the case of identical frequencies
\cite{Reddy98,Reddy99}. In these works, a pair of identical Stuart-Landau
equations are considered under scalar and diffusive coupling. A linear
stability analysis and numerical simulations are used to show that the
presence of delay can stabilize the equilibrium solution in this case. A
similar conclusion is drawn for an all-to-all coupled network of identical
Stuart-Landau systems, where the coupling strengths and time delays between
pairs are also identical. From a different perspective, other studies have
shown the role of feedback delays in suppressing or modifying oscillations in
a single oscillator \cite{Atay-JSV98,Atay-IJC02,Atay-LNCIS02}. Time delays are
thus significant in the oscillations and stability of many systems, which
justifies the efforts devoted to their analysis.

The purpose of the present paper is to obtain rigorous conditions for
amplitude death when weakly nonlinear planar systems (such as those near a
Hopf bifurcation) are diffusively coupled and the coupling involves time
delays. We consider arbitrary nonlinearities and network structure to prove
the generality of amplitude death for the interconnection of such systems.
Furthermore, we show the existence of parameter ranges for which only parts of
the network are quenched while the others continue to oscillate, even though
they remain connected. This phenomenon, which may be termed partial amplitude
death, is shown to be possible in case when the individual units are
sufficiently different.

This study extends the previous results in several directions. Firstly, we
treat a general class of systems and derive conditions for delay-induced
amplitude death irrespective of the particular nonlinearities that may be
present. In the absence of delays, general conclusions can be drawn by
considering normal forms that represent the dynamics on a center manifold when
each oscillator is near a Hopf bifurcation \cite{Aronson90}. However, the
analysis is more difficult when delays are introduced, and as a result, the
study of amplitude death in the delayed case so far has been confined to some
special equations. For instance, the system studied in \cite{Reddy98,Reddy99}
is obtained by adding delays to the center manifold equations considered in
\cite{Aronson90}, and as such it does not represent general center manifold
dynamics for delay-coupled oscillators. Hence, an interesting question is what
conclusions can be drawn in the presence of delays when nonlinearities other
than the Stuart-Landau system are considered.

Secondly, we consider very general coupling conditions. We do not restrict to
scalar coupling and present results for arbitrary diffusion matrices between
oscillators. Both the diffusion matrices and the delays are allowed to vary
from pair to pair within the network. In addition, there are no assumptions on
either the size or the topology of the network, and the results apply to any
network layout, including all-to-all, nearest-neighbor, and sparse coupling types.

Finally, instead of a local stability analysis we obtain global results using
averaging theory. Among other things, this allows the results to be applicable
to more general systems such as bistable oscillators, where the origin is
stable but there are also limit cycle solutions which may be annihilated by
the effect of coupling. In fact, in such cases the local stability analysis of
the origin cannot yield parameter ranges for amplitude death since the origin
always remains stable. On the other hand, the averaging approach enables one
to study the (averaged) trajectories inside a large sphere and derive
conditions that ensure that they all decay to zero. Furthermore, this
non-local view is essential for investigating the phenomenon partial amplitude
death, since here also, the stability type of the origin for the overall
system remains unchanged as parts of the network collapse to approach the zero solution.

The paper is organized as follows. In the next section a pair of coupled
oscillators is considered. After a brief discussion of averaging theory for
delay-differential equations, the averaged equations are obtained, which
indicate that qualitatively there are two cases to consider. Section
\ref{sec:similar} treats the first case where the frequencies of individual
oscillators are close to each other. Here, sufficient conditions are derived
for the occurrence of amplitude death, and the possibility of death for
identical oscillators for almost all positive values of the delay is proved.
The case of large frequency differences is addressed in Section
\ref{sec:dissimilar}, where it is shown that it is possible for only one of
the oscillators to be suppressed while the other one continues to oscillate.
Section \ref{sec:network} extends the results to an arbitrary size network
with varying coupling coefficients and delays between the oscillators. A
general condition is derived under which the network experiences amplitude
death as a whole or only in parts. Results are illustrated on numerical examples.

For notation, $\left\|  \cdot\right\|  $ is used for the Euclidean norm. If
$K$ is a matrix, then $K^{T}$ denotes its transpose, $\operatorname*{tr}K$ its
trace, and $K_{s}=\frac{1}{2}(K^{T}+K)$ its symmetric part. An identity matrix
of appropriate size is denoted by $I$, while $J$ denotes the skew-symmetric
matrix%
\[
J=\left[
\begin{array}
[c]{cc}%
0 & 1\\
-1 & 0
\end{array}
\right]  .
\]
Furthermore, the derivative of a function $f$ is written as $Df$.

\section{A pair of coupled oscillators}

We consider oscillators that can be modelled by equations of the form%
\begin{equation}
\dot{x}_{i}(t)=A_{i}x_{i}(t)+\varepsilon f_{i}(x_{i}(t))\,, \label{single}%
\end{equation}
where $x_{i}\in\mathbf{R}^{2}$, the matrix $A_{i}\in\mathbf{R}^{2\times2}$ has
a pair of purely imaginary eigenvalues $\pm\mathrm{i}\omega_{i}\neq0$, and
$\varepsilon$ is a nonnegative parameter. The function $f_{i}$ includes any
nonlinear terms and is assumed to satisfy the following hypotheses.

\begin{description}
\item [(H)]The function $f_{i}:\mathbf{R}^{2}\rightarrow\mathbf{R}^{2}$ is
$C^{2}$ and $f_{i}(0)=0$.
\end{description}

It is further assumed that $A_{i}$ is put into the real Jordan form%
\begin{equation}
A_{i}=\omega_{i}J=\left[
\begin{array}
[c]{cc}%
0 & \omega_{i}\\
-\omega_{i} & 0
\end{array}
\right]  \label{Omega}%
\end{equation}
with $\omega_{i}>0$, possibly after a linear coordinate transformation.
Equation (\ref{single}) typically arises from a center manifold reduction of
higher order systems near a Hopf bifurcation. Alternatively, it may be viewed
as a perturbation of the harmonic oscillator, which includes the classical
pendulum, van der Pol, and Duffing equations as special cases.

Denoting $\Phi_{i}(t)=\exp(tA_{i})$, a transformation of variables given by
$x_{i}=\Phi_{i}(t)u_{i}$ puts (\ref{single}) into the form
\begin{equation}
\dot{u}_{i}=\varepsilon\Phi_{i}^{-1}(t)f_{i}(\Phi_{i}(t)u_{i}).
\label{u-single}%
\end{equation}
This time-dependent equation can be made autonomous by averaging theory. The
averaged equation corresponding to (\ref{u-single}) is defined as
\begin{equation}
\dot{u}_{i}=\varepsilon\bar{f}_{i}(u_{i})\,, \label{u-avg}%
\end{equation}
where%
\begin{equation}
\bar{f}_{i}(u)\,=\lim_{T\rightarrow\infty}\frac{1}{T}\int_{0}^{T}\Phi_{i}%
^{-1}(t)f_{i}(\Phi_{i}(t)u)\,dt\,. \label{f-avg}%
\end{equation}
By the averaging theorem \cite{Guckenheimer-Holmes}, nonzero and hyperbolic
equilibria of (\ref{u-avg}) correspond to periodic solutions of (\ref{single})
for all sufficiently small $\varepsilon$. If one considers limit cycle
oscillators, (\ref{u-avg}) will usually have one or more asymptotically stable
equilibrium points and the origin will be unstable; however, these assumptions
are not essential for the purposes of the present paper.

Now suppose a pair of systems of the form (\ref{single}) are coupled
diffusively,%
\begin{equation}%
\begin{array}
[c]{c}%
\dot{x}_{1}(t)=A_{1}x_{1}(t)+\varepsilon f_{1}(x_{1}(t))+\varepsilon
K(x_{2}(t-\tau)-x_{1}(t))\\
\dot{x}_{2}(t)=A_{2}x_{1}(t)+\varepsilon f_{2}(x_{2}(t))+\varepsilon
K(x_{1}(t-\tau)-x_{2}(t))\,
\end{array}
\label{coupled}%
\end{equation}
Here $K\in\mathbf{R}^{2\times2}$ is a matrix of coupling coefficients and
$\tau$ is a nonnegative number representing the transmission delay. The matrix
$K$ is scaled by the parameter $\varepsilon$ to indicate that the coupling
strength is assumed to be of the same order of magnitude as the attraction to
the limit cycles of individual oscillators. Following the change of variables
$x_{i}=\Phi_{i}(t)u_{i}$, (\ref{coupled}) becomes%
\begin{equation}%
\begin{array}
[c]{c}%
\dot{u}_{1}(t)=\varepsilon\Phi_{1}^{-1}(t)f_{1}(\Phi_{1}(t)u_{1}%
(t))+\varepsilon\Phi_{1}^{-1}(t)K\left(  \Phi_{2}(t-\tau)u_{2}(t-\tau
)-\Phi_{1}(t)u_{1}(t)\right) \\
\dot{u}_{2}(t)=\varepsilon\Phi_{2}^{-1}(t)f_{2}(\Phi_{2}(t)u_{2}%
(t))+\varepsilon\Phi_{2}^{-1}(t)K\left(  \Phi_{1}(t-\tau)u_{1}(t-\tau
)-\Phi_{2}(t)u_{2}(t)\right)
\end{array}
\label{u-coupled}%
\end{equation}
For small $\varepsilon$, $u_{1}$ and $u_{2}$ are slowly changing variables and
the method of averaging is applicable.

Averaging theory for functional differential equations such as
(\ref{u-coupled}) is in principle similar to the theory for ordinary
differential equations. To introduce some notation, let $y$ denote the pair
$(u_{1},u_{2})\in\mathbf{R}^{n}$, with $n=4$ in the present case. Then
(\ref{u-coupled}) is a relation between the derivative $\dot{y}(t)$ and the
present and past values of $y(t)$. A solution $y(t)$ of (\ref{u-coupled})
describes a trajectory on the state space $\mathcal{C}=C([-\tau,0],\mathbf{R}%
^{n})$, the space of continuous functions mapping the interval $[-\tau,0]$ to
$\mathbf{R}^{n}$. The trajectory consists of points $y_{t}$ which are defined
by $y_{t}(\theta)=y(t+\theta)$, $\theta\in\lbrack-\tau,0]$. Hence each point
$y_{t}\in\mathcal{C}$ on the trajectory corresponds to a window of the
solution $y(t)$ over an interval of length $\tau$. The system (\ref{u-coupled}%
) is a particular case of the more general functional differential equation%
\begin{equation}
\dot{y}(t)=\varepsilon g(t,y_{t})\,, \label{FDE}%
\end{equation}
where $y\in\mathbf{R}^{n}$ and $g:\mathbf{(R}\times\mathcal{C})\rightarrow
\mathbf{R}^{n}$. For small values $\varepsilon$, (\ref{FDE}) can be viewed as
a perturbation of a trivial ordinary differential equation. In fact, if the
following limit exists%
\[
\bar{g}(\varphi)=\lim_{T\rightarrow\infty}\frac{1}{T}\int_{0}^{T}%
g(t,\varphi)\,dt\,,
\]
then the averaged equation corresponding to (\ref{FDE}) is defined to be
\[
\dot{y}(t)=\varepsilon\bar{g}(y_{t})
\]
in which $y_{t}$ is interpreted as a constant element of $\mathcal{C}$. In
this way, averaging reduces the infinite-dimensional system (\ref{FDE}) to an
ordinary differential equation on $\mathbf{R}^{n}$ \cite{Hale66}. In the
following, we shall exploit this fact for the analysis of (\ref{u-coupled}).
The main averaging result is contained in next lemma.

\begin{lemma}
\label{thm:Kbar} Let $\Phi_{i}=\exp(tA_{i})$, where $A_{i}$ is given by
(\ref{Omega}), and let $K\in\mathbf{R}^{2\times2}$. Then
\begin{equation}
\lim_{T\rightarrow\infty}\frac{1}{T}\int_{0}^{T}\Phi_{i}^{-1}(t)K\Phi
_{j}(t-\tau)\,dt=\left\{
\begin{array}
[c]{ccc}%
\bar{K}\Phi_{j}(-\tau) & \text{if} & \omega_{i}=\omega_{j}\\
0 & \text{if} & \omega_{i}\neq\omega_{j}%
\end{array}
\right.  \label{basic-avg}%
\end{equation}
where
\begin{equation}
\bar{K}=\frac{1}{2}(\operatorname*{tr}K)\cdot I-\frac{1}{2}\operatorname*{tr}%
(JK)\cdot J.\label{Kbar}%
\end{equation}
\end{lemma}

\begin{proof}
By (\ref{Omega}) one has%
\begin{equation}
\Phi_{i}(t)=\exp(tA_{i})=\left[
\begin{array}
[c]{cc}%
\cos\omega_{i}t & \sin\omega_{i}t\\
-\sin\omega_{i}t & \cos\omega_{i}t
\end{array}
\right]  .\label{fund-matrix}%
\end{equation}
Clearly, $\Phi_{i}^{-1}(t)=\Phi_{i}^{T}(t)$ and $\Phi_{i}(t-\tau)=\Phi
_{i}(t)\Phi_{i}(-\tau)$. Thus, if $L$ denotes the left hand side of
(\ref{basic-avg}),
\begin{equation}
L=\left[  \lim_{T\rightarrow\infty}\frac{1}{T}\int_{0}^{T}\Phi_{i}^{T}%
(t)K\Phi_{j}(t)\,dt\right]  \Phi_{j}(-\tau).\label{qq65}%
\end{equation}
When $\omega_{i}=\omega_{j}=\omega$, the integrand above consists of second
order homogenous polynomials in the variables $\sin\omega t$ and $\cos\omega
t$. Let $[k_{mn}]$ denote the elements of $K$. Then, using the elementary
facts that
\[
\lim_{T\rightarrow\infty}\frac{1}{T}\int_{0}^{T}\sin^{2}\omega t\,dt=\lim
_{T\rightarrow\infty}\frac{1}{T}\int_{0}^{T}\cos^{2}\omega t\,dt=\frac{1}{2}%
\]
and%
\[
\lim_{T\rightarrow\infty}\frac{1}{T}\int_{0}^{T}\sin\omega t\cos\omega
t\,dt=0,
\]
one can calculate%
\begin{align*}
\lim_{T\rightarrow\infty}\frac{1}{T}\int_{0}^{T}\Phi_{i}^{T}(t)K\Phi
_{j}(t)\,dt &  =\frac{1}{2}\left[
\begin{array}
[c]{cc}%
k_{11}+k_{22} & k_{12}-k_{21}\\
k_{21}-k_{12} & k_{11}+k_{22}%
\end{array}
\right]  \\
&  =\frac{1}{2}(\operatorname*{tr}K)\cdot I-\frac{1}{2}\operatorname*{tr}%
(JK)\cdot J.
\end{align*}
So, $L=\bar{K}\Phi_{j}(-\tau)$ for the case of equal frequencies. On the other
hand if $\omega_{i}\neq\omega_{j}$, then the entries of the matrix $\Phi
_{i}^{T}(t)K\Phi_{j}(t)$ are linear combinations of terms of the form
$p(\omega_{i}t)q(\omega_{j}t)$, where $p,q$ can be the sine or the cosine
functions. It follows by the theory of almost periodic functions \cite{Bohr}
that the average in the above sense of such terms is zero whenever $\omega
_{i}\neq\omega_{j}$.
\end{proof}

As the lemma implies, it is necessary to distinguish between two qualitatively
different cases, determined by the difference $|\omega_{1}-\omega_{2}|$
between the frequencies. These cases shall be addressed in the next two sections.

\section{Similar oscillators and amplitude death}

\label{sec:similar} We first consider the coupling of two oscillators with
similar native frequencies. More precisely, suppose that $|\omega_{1}%
-\omega_{2}|=\mathcal{O}(\varepsilon)$. This includes the particular case of
identical frequencies. In fact, writing $A_{1}=A_{2}+$ $\mathcal{O}%
(\varepsilon)$, and subsuming the $\mathcal{O}(\varepsilon)$ difference into
the definition of the function $f_{1}$ or $f_{2}$ in (\ref{coupled}), it is
clear that there is no loss of generality in taking $\omega_{1}=\omega_{2}$,
which will be assumed for the rest of this section.

With $\omega_{1}=\omega_{2}=\omega$, applying Lemma \ref{thm:Kbar} to the
coupled system (\ref{u-coupled}) gives the averaged equations%
\begin{equation}%
\begin{array}
[c]{c}%
\dot{u}_{1}=\varepsilon(\bar{f}_{1}(u_{1})-\bar{K}u_{1}+\bar{K}\Phi
(-\tau)u_{2})\\
\dot{u}_{2}=\varepsilon(\bar{f}_{2}(u_{2})-\bar{K}u_{2}+\bar{K}\Phi
(-\tau)u_{1})
\end{array}
\label{u-coupled-avg}%
\end{equation}
where $\Phi=\Phi_{1}=\Phi_{2}$ and the $\bar{f}_{i}$ are defined by
(\ref{f-avg}). We shall show the existence of parameter values such that
inside an arbitrary ball centered at the origin all solutions $(u_{1}%
(t),u_{2}(t))$ of (\ref{u-coupled-avg}) approach zero as $t\rightarrow\infty$.
This will imply the amplitude death in the original equations (\ref{coupled})
for sufficiently small $\varepsilon$.

\begin{theorem}
\label{thm:death} Suppose $K\in\mathbf{R}^{2\times2}$ and $R>0$,$\ $and let
the numbers $q_{i}$, $i=1,2$, be defined by%
\begin{equation}
q_{i}=\sup\left\{  \frac{u^{T}\bar{f}_{i}(u)}{\left\|  u\right\|  ^{2}%
}:\left\|  u\right\|  \leq R,\;u\neq0\right\}  .\label{Q}%
\end{equation}
If
\begin{equation}
\operatorname*{tr}K-\left|  (\operatorname*{tr}K)\cos\omega\tau
-\operatorname*{tr}(JK)\sin\omega\tau\right|  >2\max\{q_{1},q_{2}%
\}\label{stability1}%
\end{equation}
then all solutions of (\ref{u-coupled-avg}) satisfy $\lim_{t\rightarrow\infty
}(u_{1}(t),u_{2}(t))=0$ whenever $\left\|  (u_{1}(0),u_{2}(0))\right\|  \leq
R$. On the other hand, if%
\begin{equation}
\operatorname*{tr}K<\frac{1}{2}\operatorname*{tr}(Df_{1}(0)+Df_{2}%
(0))\label{instability1}%
\end{equation}
then the zero solution of (\ref{u-coupled-avg}) is unstable.
\end{theorem}

\begin{proof}
Under the conditions (H), it is easy to see that the supremum in (\ref{Q}) is
finite. In fact, since the integrand in (\ref{f-avg}) is periodic in $t$,
\begin{equation}
\bar{f}_{i}(u)\,=\frac{1}{T^{\prime}}\int_{0}^{T^{\prime}}\Phi_{i}%
^{-1}(t)f_{i}(\Phi_{i}(t)u)\,dt\label{f-avg2}%
\end{equation}
where $T^{\prime}$ is the period. Therefore the $\bar{f}_{i}$ are
differentiable and vanish at the origin, and one has the finite Taylor series
expansion%
\begin{equation}
\bar{f}_{i}(u)=D\bar{f}_{i}(0)u+\mathcal{R}_{i}(u)\label{f-Taylor}%
\end{equation}
where the remainder term $\mathcal{R}_{i}$ is continuous and $\lim
_{u\rightarrow0}\left\|  \mathcal{R}_{i}(u)\right\|  /\left\|  u\right\|
=0$.$\ $Hence $u^{T}\bar{f}_{i}(u)/\left\|  u\right\|  ^{2}$ is bounded inside
any ball with the origin removed, so the $q_{i}$ defined by (\ref{Q}) are
finite. Now taking the inner product of the first equation in
(\ref{u-coupled-avg}) by $u_{1}$ and the second one by $u_{2}$, and adding,
one obtains%
\begin{align}
\frac{1}{2}\frac{d}{dt}(\left\|  u_{1}\right\|  ^{2}+\left\|  u_{2}\right\|
^{2}) &  =\varepsilon\left(  u_{1}^{T}\bar{f}_{1}(u_{1})+u_{2}^{T}\bar{f}%
_{2}(u_{2})-u_{1}^{T}\bar{K}u_{1}-u_{2}^{T}\bar{K}u_{2}\right)  \nonumber\\
&  +\varepsilon\left(  u_{1}^{T}\bar{K}\Phi(-\tau)u_{2}+u_{2}^{T}\bar{K}%
\Phi(-\tau)u_{1}\right)  .\label{qq10}%
\end{align}
By (\ref{Q}) $u_{i}^{T}\bar{f}_{i}(u_{i})\leq q_{i}\left\|  u_{i}\right\|
^{2}$. Hence, if $v=(u_{1},u_{2})$ denotes the vector formed by concatenating
$u_{1}$ and $u_{2}$, then (\ref{qq10}) yields~%
\begin{equation}
\frac{d}{dt}\left\|  v\right\|  ^{2}\leq2\varepsilon v^{T}Pv\label{qq21}%
\end{equation}
where%
\begin{equation}
P=\left[
\begin{array}
[c]{cc}%
q_{1}I-\bar{K} & \bar{K}\Phi(-\tau)\\
\bar{K}\Phi(-\tau) & q_{2}I-\bar{K}%
\end{array}
\right]  \text{.}\label{P}%
\end{equation}
Clearly, $v^{T}Pv=v^{T}P_{s}v$, where $P_{s}$ denotes the symmetric part of
$P$. We claim that $P_{s}$ is negative definite if the condition
(\ref{stability1}) is satisfied. Indeed, it is immediate from (\ref{Kbar})
that $\bar{K}_{s}=$ $\frac{1}{2}(\operatorname*{tr}K)I$, and a calculation
using (\ref{fund-matrix}) shows that
\begin{equation}
(\bar{K}\Phi(-\tau))_{s}=\frac{1}{2}[(\operatorname*{tr}K)\cos\omega
\tau-\operatorname*{tr}(JK)\sin\omega\tau]\,I.\label{Kbar-sym2}%
\end{equation}
Hence, $P_{s}$ has the form%
\begin{equation}
P_{s}=\left[
\begin{array}
[c]{cc}%
q_{1}I-\bar{K}_{s} & \left(  \bar{K}\Phi(-\tau)\right)  _{s}\\
\left(  \bar{K}\Phi(-\tau)\right)  _{s} & q_{2}I-\bar{K}_{s}%
\end{array}
\right]  =\frac{1}{2}\left[
\begin{array}
[c]{cc}%
\eta_{1}I & \alpha I\\
\alpha I & \eta_{2}I
\end{array}
\right]  \label{qq61}%
\end{equation}
with
\[
\eta_{i}=2q_{i}-\operatorname*{tr}K,\quad i=1,2
\]
and
\[
\alpha=(\operatorname*{tr}K)\cos\omega\tau-\operatorname*{tr}(JK)\sin
\omega\tau.
\]
If (\ref{stability1}) holds, then $\eta_{1}$ and $\eta_{2}$ are both negative
and $\eta_{1}\eta_{2}>\alpha^{2}$. Thus, by the well-known tests for
definiteness, the matrix $P_{s}$ in (\ref{qq61}) is negative definite, and the
claim is proved. It now follows under these conditions that
\[
v^{T}Pv=v^{T}P_{s}v\leq-\lambda\left\|  v\right\|  ^{2}\quad\text{for all
}v\in\mathbf{R}^{4}%
\]
where $-\lambda<0$ is the largest eigenvalue of $P_{s}$. Substituting into
(\ref{qq21}) gives
\[
\left\|  v(t)\right\|  ^{2}\leq\left\|  v(0)\right\|  ^{2}\exp(-2\varepsilon
\lambda t)\quad\text{for }t\geq0
\]
which establishes the first statement of the theorem.

To prove the second statement, consider the linearization of
(\ref{u-coupled-avg}) about zero:%
\begin{equation}
\dot{v}=\varepsilon\left[
\begin{array}
[c]{cc}%
D\bar{f}_{1}(0)-\bar{K} & \bar{K}\Phi(-\tau)\\
\bar{K}\Phi(-\tau) & D\bar{f}_{2}(0)-\bar{K}%
\end{array}
\right]  v\label{qq22}%
\end{equation}
By (\ref{f-avg2}) and Lemma \ref{thm:Kbar},%
\begin{align}
D\bar{f}_{i}(0)\,=\frac{1}{T^{\prime}}\int_{0}^{T^{\prime}}\Phi_{i}%
^{-1}(t)Df_{i}(0)\Phi_{i}(t)\,dt\nonumber\\
=\frac{1}{2}\operatorname*{tr}[Df_{i}(0)]\cdot I-\frac{1}{2}\operatorname*{tr}%
(J[Df_{i}(0)])\cdot J.\label{qq24}%
\end{align}
Since $J$ has zero trace,
\begin{equation}
\operatorname*{tr}D\bar{f}_{i}(0)=\operatorname*{tr}Df_{i}(0).\label{qq25}%
\end{equation}
Also, $\operatorname*{tr}K=\operatorname*{tr}\bar{K}$ by (\ref{Kbar}). Thus
the matrix in (\ref{qq22}) has trace equal to
\[
\operatorname*{tr}(Df_{1}(0)+Df_{2}(0)-2K)
\]
which is positive if (\ref{instability1}) holds, in which case at least one of
its eigenvalues has a positive real part.
\end{proof}

\begin{description}
\item [ Remark. ]The quantities $q_{i}$ defined in (\ref{Q}) satisfy the lower
bounds%
\begin{equation}
q_{i}\geq\frac{1}{2}\operatorname*{tr}[Df_{i}(0)],\label{q-estimate}%
\end{equation}
which provides a connection between the stability condition (\ref{stability1})
and the instability condition (\ref{instability1}). The inequality
(\ref{q-estimate}) follows by taking limits in (\ref{f-Taylor}) and using
(\ref{qq24}) to obtain%
\[
\lim_{u\rightarrow0}\frac{u^{T}\bar{f}_{i}(u)}{\left\|  u\right\|  ^{2}}%
=\lim_{u\rightarrow0}\frac{u^{T}[D\bar{f}_{i}(0)]u}{\left\|  u\right\|  ^{2}%
}=\frac{1}{2}\operatorname*{tr}[Df_{i}(0)].
\]
Thus if $q_{i}$ satisfies (\ref{Q}) for any $R>0$ then necessarily $q_{i}%
\geq\frac{1}{2}$ $\operatorname*{tr}[Df_{i}(0)]$. This relates $q_{i}$ to the
local stability of the origin for the individual uncoupled oscillator. For
instance, if the linear part of the oscillator (\ref{single}) has the
eigenvalues $\alpha\pm\mathrm{i}\omega$, then $q_{i}\geq\alpha$ by
(\ref{q-estimate}). In particular, $q_{i}>0$ in case the oscillations arise
from a supercritical Hopf bifurcation. On the other hand, if the origin is
asymptotically stable then $\operatorname*{tr}Df_{i}(0)<0$, and $q_{i}$ may be
negative in a sufficiently small neighborhood of the origin, but not
necessarily in larger neighborhoods (see Example~1 below). To avoid trivial
cases from the point of study of the death of limit cycles, it will be assumed
in this paper that the $q_{i}$ are positive in the region of interest.
\end{description}

\begin{figure}[tb]
\begin{center}
\includegraphics{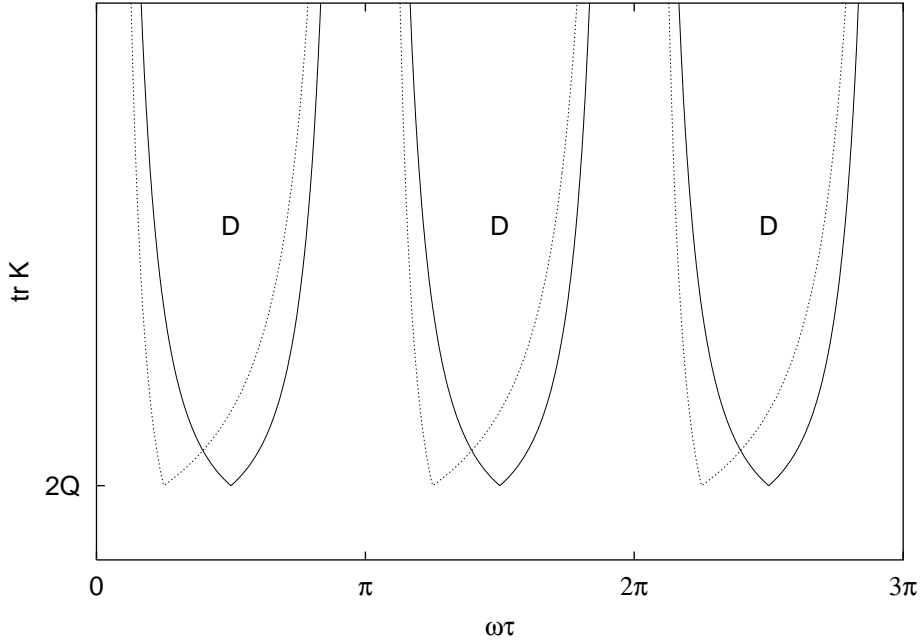}
\end{center}
\caption{Regions marked ``D'' are parameter values for which amplitude death
occurs, and $Q=\max\{q_{1},q_{2}\}$. The solid lines correspond to the
case when the matrix $K$ is symmetric. For a nonsymmetric
$K$, the death regions are warped to the side, as shown by dotted lines.}%
\label{fig:death}%
\end{figure}

In essence, Theorem \ref{thm:death} states that in case of equal frequencies,
amplitude death occurs whenever the trace of the matrix $K$ is sufficiently
large, provided that $\sin\omega\tau\neq0$. By the averaging theory, for a set
of parameter values satisfying (\ref{stability1}) there exists $\varepsilon
_{0}>0$ such that the original coupled system (\ref{coupled}) undergoes
amplitude death if $\varepsilon\in(0,\varepsilon_{0})$. Condition
(\ref{stability1}) is graphically depicted on the parameter plane in Figure
\ref{fig:death}. It should be noted that the graph is not uniform in
$\varepsilon$. For instance, it may be necessary to decrease $\varepsilon$ as
$\operatorname*{tr}K$ increases in order to observe amplitude death. This
should be expected since the foregoing analysis assumes that the coupling
coefficients $\varepsilon K$ are $\mathcal{O}(\varepsilon)$ terms.
Nevertheless, the value of $\varepsilon_{0}$ can be chosen to be constant over
any compact subset of the parameter space. The extent of the death region in
the figure suggests that amplitude death is quite a robust dynamical behavior
for diffusively coupled systems.

Some general qualitative characteristics of the region of amplitude death can
be deduced from Figure~\ref{fig:death}. For instance, the death region is a
disconnected set, formed of ``islands of death'' as mentioned in
\cite{Reddy99}. Furthermore, it is bounded away from the horizontal axis,
implying that there is a threshold of coupling strength below which amplitude
death is not possible. Strictly speaking, these properties do not follow from
(\ref{stability1}), which is only a sufficient condition for amplitude death.
Nevertheless, they can be established directly by a linear stability analysis
of the origin. Thus, the statement that $\operatorname*{tr}K$ should be
sufficiently large for amplitude death follows from the instability condition
(\ref{instability1}). Similarly, it can be shown that amplitude does not occur
if $\omega\tau=n\pi$ for some integer $n$. Indeed, in this case $\Phi
(-\tau)=(-1)^{n}I$ by (\ref{fund-matrix}), and the linear variational equation
(\ref{qq22}) takes the form
\[
\dot{v}=\varepsilon\left[
\begin{array}
[c]{cc}%
D\bar{f}(0)-\bar{K} & (-1)^{n}\bar{K}\\
(-1)^{n}\bar{K} & D\bar{f}(0)-\bar{K}%
\end{array}
\right]  v
\]
where for simplicity it is assumed that the oscillators are identical, and
$f=f_{1}=f_{2}$. Letting $C$ denote the matrix on the right side, we note that%
\[
C\left[
\begin{array}
[c]{c}%
I\\
(-1)^{n}I
\end{array}
\right]  =\left[
\begin{array}
[c]{c}%
I\\
(-1)^{n}I
\end{array}
\right]  D\bar{f}(0).
\]
If $\lambda$ is an eigenvalue of $D\bar{f}(0)$ corresponding to an eigenvector
$p$, then%
\[
C\left[
\begin{array}
[c]{c}%
I\\
(-1)^{n}I
\end{array}
\right]  p=\left[
\begin{array}
[c]{c}%
I\\
(-1)^{n}I
\end{array}
\right]  [D\bar{f}(0)]p=\lambda\left[
\begin{array}
[c]{c}%
I\\
(-1)^{n}I
\end{array}
\right]  p
\]
i.e., $\lambda$ is also an eigenvalue of $C$. From (\ref{qq24}) it is seen
that $\frac{1}{2}\operatorname*{tr}[Df(0)]\pm\mathrm{i}\frac{1}{2}%
\operatorname*{tr}(J[Df(0)])$ are the eigenvalues of $D\bar{f}(0)$, and thus
of $C$. Consequently, the origin is unstable if $\operatorname*{tr}[Df(0)]>0$,
e.g. when the oscillations result from a supercritical Hopf bifurcation, as
noted in the Remark above. This proves that amplitude death is not possible
for identical oscillators if $\omega\tau=n\pi,$ $n\in\mathbf{Z}$; in
particular death does not occur for $\tau=0$, which agrees with earlier
results obtained for the undelayed case \cite{Aronson90}. It should be pointed
out that a disconnected death region in the parameter space arises under the
assumptions that $\varepsilon$ is small and the delays are discrete. For large
values of $\varepsilon$ single or multiple death islands (or none) may be
possible, as was shown for the delayed Stuart-Landau equations in
\cite{Reddy99}. On the other hand, recent results indicate that regions of
amplitude death typically grow and merge if the delays are distributed over an
interval rather than concentrated at a single point \cite{Atay-MPI03a}.

The global nature of Theorem \ref{thm:death} makes it applicable also to cases
where a linear stability analysis may be inconclusive---for instance, to
bistable oscillators, as the next example illustrates.

\begin{description}
\item [ Example 1 ]Consider the following variants of van der Pol oscillators
which are coupled through their velocities:
\begin{equation}%
\begin{array}
[c]{c}%
\ddot{z}_{1}+\varepsilon\dot{z}_{1}(1-(1+\rho^{2})z_{1}^{2}+\frac{1}{2}%
\rho^{2}z_{1}^{4})+z_{1}=\varepsilon\beta(\dot{z}_{2}(t-\pi/2)-\dot{z}_{1}%
(t))\\
\ddot{z}_{2}+\varepsilon\dot{z}_{2}(1-(1+\rho^{2})z_{2}^{2}+\frac{1}{2}%
\rho^{2}z_{2}^{4})+z_{2}=\varepsilon\beta(\dot{z}_{1}(t-\pi/2)-\dot{z}_{2}(t))
\end{array}
\label{bistable}%
\end{equation}
For $\rho=0$ and $\varepsilon<0$ the left hand sides of (\ref{bistable})
describe the usual van der Pol oscillators. The scalar $\beta$ denotes the
coupling strength. Putting (\ref{bistable}) into the vector form
(\ref{coupled}) by letting $x_{i}=(z_{i},\dot{z}_{i})$, it is seen that%
\[
K=\left[
\begin{array}
[c]{cc}%
0 & 0\\
0 & \beta
\end{array}
\right]  ,
\]
thus $\operatorname*{tr}K=\beta$. In the absence of coupling, the averaged
equations have the form (\ref{u-avg}) with%
\begin{equation}
\bar{f}_{i}(u)=-\frac{1}{2}\left(  1-\frac{(1+\rho^{2})}{4}\left\|  u\right\|
^{2}+\frac{\rho^{2}}{16}\left\|  u\right\|  ^{4}\right)  \cdot u,\quad
i=1,2.\label{qq15}%
\end{equation}
The roots $u$ of $\bar{f}_{i}$ satisfy $\left\|  u\right\|  =0,2,$ or $2/\rho
$. For definiteness we take $\rho=2$. Then for $\beta=0$ and $\varepsilon$
small and positive, each (uncoupled) oscillator has two attractors: the zero
solution and a periodic solution with amplitude near 2. In addition, there is
an unstable periodic solution whose amplitude is near 1. Now by (\ref{qq15}),%
\begin{equation}
u^{T}\bar{f}_{i}(u)=-\frac{1}{2}\left(  1-\frac{5}{4}\left\|  u\right\|
^{2}+\frac{1}{4}\left\|  u\right\|  ^{4}\right)  \cdot\left\|  u\right\|
^{2}\label{qq16}%
\end{equation}
from which, by simple calculus on the quantity in parenthesis, it follows that%
\[
u^{T}\bar{f}_{i}(u)\leq\frac{9}{32}\left\|  u\right\|  ^{2},\quad
u\in\mathbf{R}^{2},\;i=1,2
\]
so $q_{1}=q_{2}=9/32$. Theorem \ref{thm:death} then implies that for
$\beta>9/16$ and sufficiently small positive $\varepsilon$ the coupled system
has no periodic solutions and the origin is the only attractor. The agreement
with the numerical solutions of (\ref{bistable}) is shown in Figure
\ref{fig:stable} for several values of $\varepsilon$.
\begin{figure}[tb]
\begin{center}
\includegraphics{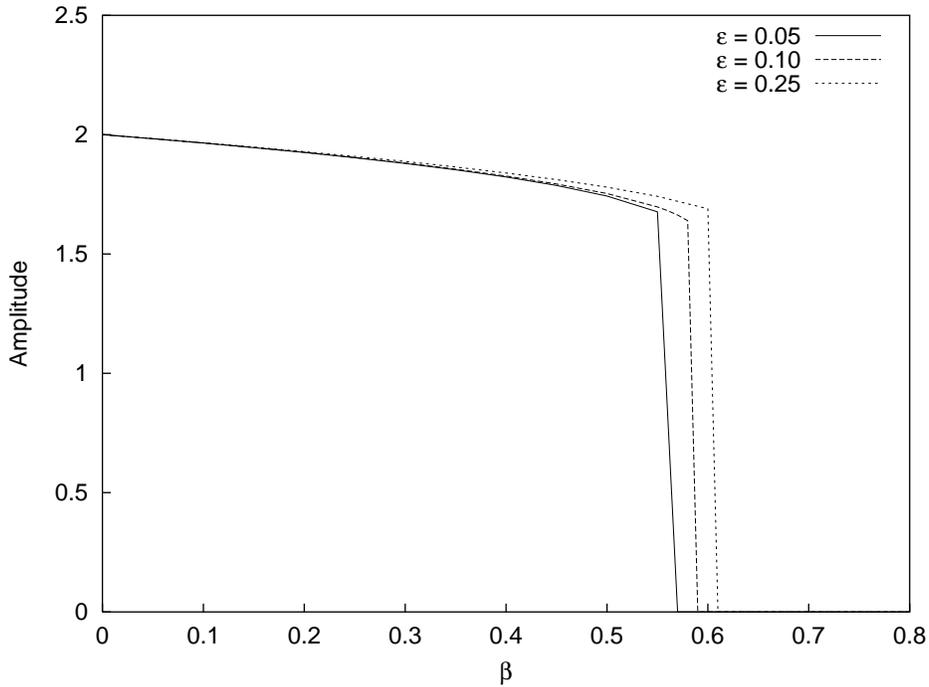}
\end{center}
\caption{Death of the periodic solutions of coupled bistable oscillators as
the coupling strength $\beta$ is varied. The zero solution remains
asymptotically stable throughout the parameter scale.}%
\label{fig:stable}%
\end{figure}
Here the vertical scale
is the amplitude (in the variables $z_{1}$ and $z_{2}$) of the attracting
periodic solution, which exists for the uncoupled system, and thus also for
small values of $\beta$. As the coupling strength $\beta$ increases past a
critical value, the periodic solutions are annihilated. The critical coupling
strength approaches $9/16\approx0.\,\allowbreak56$ as $\varepsilon$ is made
smaller, as predicted by averaging theory. A closer inspection of the averaged
equations reveals the mechanism through which amplitude death occurs. With
$\tau=\pi/2$, Lemma \ref{thm:Kbar} gives $\bar{K}=\frac{1}{2}\beta I$ and
$\bar{K}\Phi(-\tau)=-\frac{1}{2}\beta J$. The latter being an antisymmetric
matrix, the last two terms in (\ref{qq10}) vanish, leaving%
\[
\frac{1}{2}\frac{d}{dt}(\left\|  u_{1}\right\|  ^{2}+\left\|  u_{2}\right\|
^{2})=\varepsilon\left(  u_{1}^{T}\bar{f}_{1}(u_{1})+u_{2}^{T}\bar{f}%
_{2}(u_{2})-\frac{1}{2}\beta(\left\|  u_{1}\right\|  ^{2}+\left\|
u_{2}\right\|  ^{2})\right).
\]
Since the oscillators are identical and the coupling is symmetric, one could
look for equal amplitude solutions, say $\left\|  u_{1}\right\|  ^{2}=\left\|
u_{2}\right\|  ^{2}=r^{2}$. Using (\ref{qq16}) in the above equation then
gives%
\begin{equation}
\dot{r}=-\frac{\varepsilon}{2}\left(  \beta+1-\frac{5}{4}r^{2}+\frac{1}%
{4}r^{4}\right)  r\overset{def}{=}\frac{\varepsilon}{2}h(r).\label{r}%
\end{equation}
The equilibrium solutions of (\ref{r}) are $r=0$ and $r=\sqrt{\frac{5}{2}%
\pm\frac{1}{2}\sqrt{9-16\beta}}$, whenever the radicands are nonnegative. Thus
for $-1<\beta<9/16$ there are two positive equilibria corresponding to the
unstable and stable periodic solutions of the oscillators. As $\beta
\rightarrow9/16^{-}$, these two equilibria approach each other and disappear
in a saddle-node bifurcation. Correspondingly, for $\beta>9/16$ there are no
periodic solutions. The situation is graphically shown in Figure~\ref{fig:h}.
It is seen that the sufficient condition $\beta>9/16$ given by Theorem
\ref{thm:death} is also necessary for amplitude death in the present case.
Furthermore, throughout the whole range of $\beta$ values the origin remains
asymptotically stable. Hence, the local stability analysis of the origin
reveals no information about the death of the periodic solutions in this example.

\begin{figure}[tb]
\begin{center}
\includegraphics{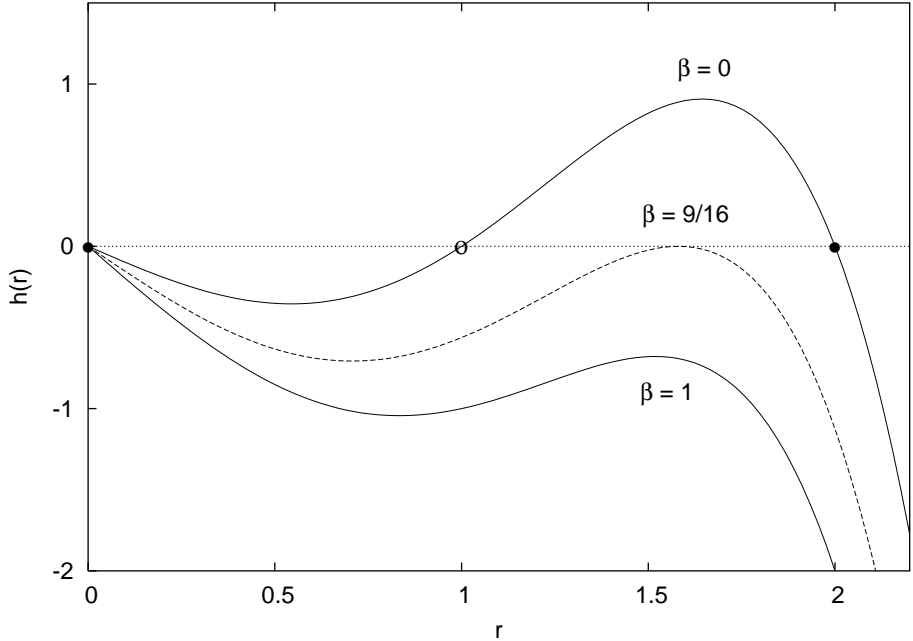}
\end{center}
\caption{The shape of the right hand side of (\ref{r}) for various values of
$\beta$. The roots correspond to the amplitudes of the stable ($\bullet$) and
unstable ($\circ$) periodic solutions. As $\beta$
is increased the two positive roots coalesce and disappear. The critical value
for this saddle-node bifurcation is $\beta=9/16$.}%
\label{fig:h}%
\end{figure}
\end{description}

\section{Dissimilar oscillators and partial amplitude death}

\label{sec:dissimilar} Suppose now that $|\omega_{1}-\omega_{2}|=\mathcal{O}%
(1)$, i.e., the frequencies differ by a nonzero constant which does not depend
on the parameter $\varepsilon$. Applying Lemma \ref{thm:Kbar} to
(\ref{u-coupled}), the averaged equations are found to be%
\begin{equation}%
\begin{array}
[c]{c}%
\dot{u}_{1}=\varepsilon(\bar{f}_{1}(u_{1})-\bar{K}u_{1})\\
\dot{u}_{2}=\varepsilon(\bar{f}_{2}(u_{2})-\bar{K}u_{2})
\end{array}
\label{u-decoupled-avg}%
\end{equation}
Note that the equations are decoupled and do not depend on the delay $\tau$.
In analogy with Theorem \ref{thm:death}, the following result can be stated.

\begin{theorem}
\label{thm:partial} Let $R>0,\ $and $q_{i}$ be defined by (\ref{Q}). If
\begin{equation}
\operatorname*{tr}K>2q_{i}\label{cond-partial1}%
\end{equation}
then all solutions of the $i$-th equation in (\ref{u-decoupled-avg}) satisfy
$\lim_{t\rightarrow\infty}u_{i}(t)=0$ whenever $\left\|  u_{i}(0)\right\|
\leq R$. On the other hand, if%
\begin{equation}
\operatorname*{tr}K<\operatorname*{tr}Df_{i}(0)\label{cond-partial2}%
\end{equation}
then the zero solution of $i$-th equation in (\ref{u-decoupled-avg}) is unstable.
\end{theorem}

\begin{proof}
Taking the inner product of $u_{i}$ and the $i$-th equation in
(\ref{u-decoupled-avg}), and using (\ref{Kbar}) and (\ref{Q}), one obtains%
\begin{align*}
\frac{1}{2}\frac{d}{dt}\left\|  u_{i}\right\|  ^{2} &  \leq\varepsilon
u_{i}^{T}(q_{1}I-\bar{K})u_{i}\\
&  =\varepsilon u_{i}^{T}(q_{1}I-\bar{K})_{s}u_{i}\\
&  =\varepsilon(q_{1}-\tfrac{1}{2}\operatorname*{tr}K)\left\|  u_{i}\right\|
^{2}%
\end{align*}
from which the first statement of the theorem follows. On the other hand, the
linearized equation%
\[
\dot{u}_{1}=\varepsilon(D\bar{f}_{i}(0)-\bar{K})u_{i}%
\]
is unstable if $\operatorname*{tr}(D\bar{f}_{i}(0)-\bar{K})$ is positive. But
the latter quantity is equal to $\operatorname*{tr}(Df_{i}(0)-K)$ as observed
in the proof of Theorem \ref{thm:death}, which establishes the second statement.
\end{proof}

Similar to Theorem \ref{thm:death}, the above result implies that the pair of
oscillators undergo amplitude death if $\operatorname*{tr}K$ is sufficiently
large, but this time regardless of the value of the delay. Comparison of the
conditions (\ref{stability1}) and (\ref{cond-partial1}) shows that amplitude
death becomes more likely when the coupled oscillators have a frequency
difference, which agrees with previous studies of amplitude death caused by
frequency distributions \cite{Ermentrout90,Mirollo90}. More importantly, in
this case the conditions for amplitude death are separate for each equation.
Thus, it is possible that one of the oscillators is suppressed independently
of the other one although they remain connected, leading to a partial
amplitude death in the coupled system. The following example illustrates this behavior.

\begin{figure}[tb]
\begin{center}
\includegraphics{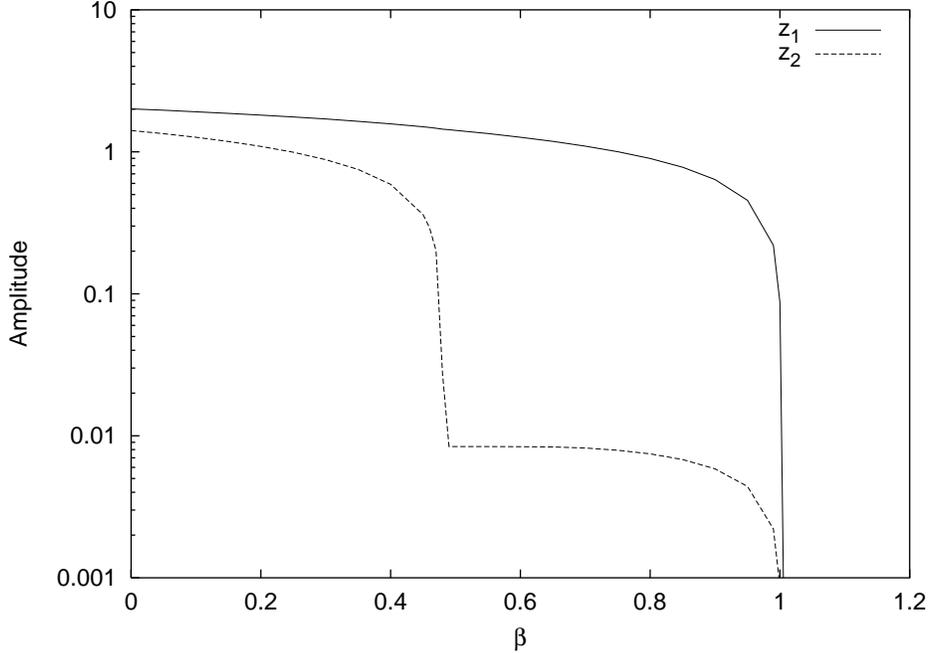}
\end{center}
\caption{Behavior of coupled van der Pol oscillators whose limit cycles have
different amplitudes and frequencies, as the coupling strength $\beta$ is
varied. Partial amplitude death occurs for $0.5<\beta<1$, where only one of
the oscillators is quenched while the other one continues to exhibit
high-amplitude oscillations. The region $\beta>1$ corresponds to the usual
amplitude death of the whole system. Note that the vertical scale is
logarithmic.}%
\label{fig:partial}%
\end{figure}

\begin{description}
\item [ Example 2 ]Consider a pair of coupled van der Pol oscillators%
\begin{equation}%
\begin{array}
[c]{c}%
\ddot{z}_{1}+\varepsilon(z_{1}^{2}-\alpha_{1})\dot{z}_{1}+\omega_{1}^{2}%
z_{1}=\varepsilon\beta(\dot{z}_{2}(t-1)-\dot{z}_{1}(t))\\
\ddot{z}_{2}+\varepsilon(z_{2}^{2}-\alpha_{2})\dot{z}_{2}+\omega_{2}^{2}%
z_{2}=\varepsilon\beta(\dot{z}_{1}(t-1)-\dot{z}_{2}(t))
\end{array}
\label{vanderpol}%
\end{equation}
where $\alpha_{i}>0$ and $\beta\in\mathbf{R}$. As in the previous example,
letting $x_{i}=(\omega_{i}z_{i},\dot{z}_{i})$ shows that $\beta$ is equal to
the trace of the matrix $K$ in (\ref{coupled}). In the averaged equations
(\ref{u-decoupled-avg}) one has%
\begin{equation}
\bar{f}_{i}(u)=\frac{1}{2}\left(  \alpha_{i}-\frac{1}{4}\left\|  u\right\|
^{2}\right)  \cdot u,\quad i=1,2\label{qq17}%
\end{equation}
which confirms that in the absence of coupling the $i$-th oscillator has an
attracting limit cycle solution $z_{i}=2\sqrt{\alpha_{i}}\cos\omega
_{i}t+\mathcal{O}(\varepsilon)$. Furthermore, (\ref{qq17}) yields%
\[
u^{T}\bar{f}_{i}(u)\leq\frac{\alpha_{i}}{2}\left\|  u\right\|  ^{2}.
\]
Hence, with $q_{i}=\alpha_{i}/2$ an application of Theorem \ref{thm:partial}
gives the condition $\beta>\alpha_{i}$ for the death of the $i$-th oscillator
in the coupled system. In particular, if $\alpha_{2}<\beta<\alpha_{1}$ then
only the second oscillator will be suppressed. The results of the numerical
solution of (\ref{vanderpol})\ is summarized in Figure \ref{fig:partial},
where the parameter values are $\alpha_{1}=1$, $\alpha_{2}=0.5$, $\omega
_{1}=0.1$, $\omega_{2}=1$, and $\varepsilon=0.1$. It is seen that amplitude
death occurs for $\beta>1$, where both oscillators are quenched. On the other
hand, a partial death occurs for $0.5<\beta<1$, where only the second
oscillator is damped while the first one continues to exhibit high-amplitude
oscillations. For this parameter range the amplitude of the second oscillator
is almost (but not exactly) zero; in fact, the amplitude approaches zero as
$\varepsilon\rightarrow0^{+}$. Indeed, the averaged equations
(\ref{u-decoupled-avg}) in this case are
\begin{equation}
\dot{u}_{i}=\frac{1}{2}\left(  \alpha_{i}-\frac{1}{4}\left\|  u_{i}\right\|
^{2}-\beta\right)  \cdot u_{i},\quad i=1,2.\label{qq19}%
\end{equation}
which have an attracting fixed point of the form $(u_{1},0)$ with $\left\|
u_{1}\right\|  =2\sqrt{\alpha_{1}-\beta}$ when $\alpha_{2}<\beta<\alpha_{1}$.
Thus for small $\varepsilon$ the amplitudes of $z_{1}$ and $z_{2}$ in the
original system (\ref{vanderpol}) are $2\sqrt{1-\beta}+\mathcal{O}%
(\varepsilon)$ and $\mathcal{O}(\varepsilon)$, respectively. It should be
remarked that, for this range of $\beta$, although the zero solution is
asymptotically stable in the equation for $u_{2}$ it is not stable for the
full system of equations (\ref{qq19}). As the origin remains unstable
throughout the range $0<\beta<1$, a linear stability analysis alone is
insufficient to reveal the occurrence of the partial amplitude death of the
coupled system. Also note that the actual value of the delay is not
significant in this behavior, and one obtains pictures similar to Figure
\ref{fig:partial} even for zero delay.
\end{description}

It is rather interesting to observe partial death under diffusive coupling,
since one intuitively expects this type of connection to act towards reducing
the differences between the oscillators, when, for instance, $K$ is a positive
scalar. For the undelayed case, results related to this behavior was given in
\cite{Atay-NDES02-conf}, which essentially follow by the observation that
weakly connected systems near a Hopf bifurcation can be decoupled in the
$\mathcal{O}(\varepsilon)$ terms if the frequencies are different, a fact
which has been applied to the study of neural networks
\cite{Hoppensteadt96,Hoppensteadt-WCNN}. Here we see that the same conclusion
also holds in the presence of arbitrary coupling delays. However, it should be
noted that higher-order terms in $\varepsilon$ may still be coupled and
dependent on $\tau$. This can be significant at large values of $\varepsilon$;
for instance, the results of \cite{Reddy99} indicate that in such a case
amplitude death may depend on the value of $\tau$, especially when the
frequency difference between the oscillators is small. In closing we mention
that a form of partial death was also observed numerically in undelayed
networks of relaxation oscillators \cite{Volkov02}. Since Theorem
\ref{thm:partial} holds under quite general conditions, partial amplitude
death can be expected to be among the common behaviors of many practical
networks formed by the interaction of dissimilar units.

\section{Network of oscillators}

\label{sec:network}The results of the previous sections will now be extended
to a general network of $N$ ~coupled oscillators modelled in the form%
\begin{equation}
\dot{x}_{i}(t)=A_{i}x_{i}(t)+\varepsilon f_{i}(x_{i}(t))+\varepsilon\sum
_{j=1}^{N}K_{ij}(x_{j}(t-\tau_{ij})-x_{i}(t))\,,\quad i=1,2,\dots,N\text{.}
\label{network}%
\end{equation}
Here $x_{i}\in\mathbf{R}^{2}$, $A_{i}=\omega_{i}J$, $f_{i}$ satisfies the
hypothesis (H), and $\varepsilon\geq0$. The nonnegative quantity $\tau_{ij}$
is the transmission delay and $K_{ij}\in\mathbf{R}^{2\times2}$ are the
coupling or diffusion coefficients between the $i$-th and the $j$-th
oscillator. To prevent self-coupling $K_{ii}$ is set to zero for all $i$. It
is assumed that the coupling between pairs is symmetric, that is,
$K_{ij}=K_{ji}$ and $\tau_{ij}=\tau_{ji}$ for all $i,j$. On the other hand,
both the coupling coefficients and the delays are allowed to vary from pair to
pair. This is an especially realistic assumption if the delays arise as a
result of spatial separation of the oscillators.

With the change of variables $x_{i}=\Phi_{i}(t)u_{i}=\exp(tA_{i})u_{i}$,
(\ref{network}) is transformed to%
\begin{equation}
\dot{u}_{i}(t)=\varepsilon\Phi_{i}^{-1}(t)f_{i}(\Phi_{i}(t)u_{i}%
(t))-\varepsilon\sum_{j=1}^{N}\Phi_{i}^{-1}(t)K_{ij}\Phi_{i}(t)u_{i}%
+\varepsilon\sum_{j=1}^{N}\Phi_{i}^{-1}(t)K_{ij}\Phi_{j}(t-\tau_{ij}%
)u_{j}(t-\tau_{ij}) \label{network-u}%
\end{equation}
for $i=1,2,\dots,N$. If $\tau=\max\{\tau_{ij}\;|\;i,j=1,\dots,N\}$, then
(\ref{network-u}) has the form (\ref{FDE}) where the right-hand side a
function on $\mathbf{R\times} C([-\tau,0],\mathbf{R}^{2N})$, and can be
averaged as before. To this end, suppose there are $M$ distinct frequencies
$\omega_{1},\dots,\omega_{M}$ among the oscillators, and let $\mathcal{I}_{n}$
denote the set of indices of those oscillators having frequency $\omega_{n}$,
for $n=1,\dots,M$. More precisely, $j\in\mathcal{I}_{n}$ if and only if the
matrix $A_{j}$ has eigenvalues $\pm\mathrm{i}\omega_{n}$. It follows by Lemma
\ref{thm:Kbar} that the average of the terms of the form $\Phi_{i}%
^{-1}(t)K_{ij}\Phi_{j}(t-\tau_{ij})$ in (\ref{network-u}) will be zero
whenever $i\in\mathcal{I}_{n}$ and $j\notin\mathcal{I}_{n}$. Application of
the lemma gives the averaged equations%
\begin{equation}
\dot{u}_{i}=\varepsilon\left(  \bar{f}_{i}(u_{i})-\sum_{j=1}^{N}\bar{K}%
_{ij}u_{i}+\sum_{j\in\mathcal{I}_{n}}\bar{K}_{ij}\Phi_{j}(-\tau_{ij}%
)u_{j}\right)  ,\quad i\in\mathcal{I}_{n}, \label{net-avg}%
\end{equation}
for $n=1,\dots,M$, where
\begin{equation}
\bar{K}_{ij}=\frac{1}{2}(\operatorname*{tr}K_{ij})\cdot I-\frac{1}%
{2}\operatorname*{tr}(JK_{ij})\cdot J\text{.} \label{Kbar-ij}%
\end{equation}
The averaged equations consist of $M$ decoupled sets of equations. The results
of the previous sections suggest that each set can undergo amplitude death
independently of the others. This observation is made precise by the following theorem.

\begin{theorem}
\label{thm:net} Suppose $K\in\mathbf{R}^{2\times2}$ and $R>0,\ $and let
$q_{i}$, $i\in\mathcal{I}_{n}$, be defined by (\ref{Q}). Let $v_{n}$ denote
the vector formed by the concatenation of $u_{i}$ for $i\in\mathcal{I}_{n}$.
If
\begin{equation}
\sum_{j=1}^{N}\operatorname*{tr}K_{ij}-\sum_{j\in\mathcal{I}_{n}}\left|
(\operatorname*{tr}K_{ij})\cos\omega_{n}\tau_{ij}-\operatorname*{tr}%
(JK_{ij})\sin\omega_{n}\tau_{ij}\right|  >2q_{i}\quad\text{for every }%
i\in\mathcal{I}_{n}\label{stab-net}%
\end{equation}
then all solutions of the $n$-th set of equations (\ref{net-avg}) satisfy
$\lim_{t\rightarrow\infty}v_{n}(t)=0$ whenever $\left\|  v_{n}(0)\right\|
\leq R$. On the other hand, if%
\begin{equation}
\sum_{i\in\mathcal{I}_{n}}\left(  \sum_{j=1}^{N}\operatorname*{tr}%
K_{ij}-\operatorname*{tr}Df_{i}(0)\right)  <0\label{instab-net}%
\end{equation}
then the zero solution of (\ref{net-avg}) is unstable.
\end{theorem}

\begin{proof}
Taking the inner product of the $i$-th equation in (\ref{net-avg}) with
$u_{i}$ and adding over $i\in\mathcal{I}_{n}$ gives%
\[
\frac{1}{2}\frac{d}{dt}\sum_{i\in\mathcal{I}_{n}}\left\|  u_{i}\right\|
^{2}=\varepsilon\sum_{i\in\mathcal{I}_{n}}\left(  u_{i}^{T}\bar{f}_{i}%
(u_{i})-\sum_{j=1}^{N}u_{i}^{T}\bar{K}_{ij}u_{i}+u_{i}^{T}\sum_{j\in
\mathcal{I}_{n}}\bar{K}_{ij}\Phi_{j}(-\tau_{ij})u_{j}\right)  .
\]
Let $R>0$. By (\ref{Kbar}) and (\ref{Q}),
\begin{equation}
\frac{d}{dt}\sum_{i\in\mathcal{I}_{n}}\left\|  u_{i}\right\|  ^{2}%
\leq2\varepsilon\sum_{i\in\mathcal{I}_{n}}\left(  q_{i}\left\|  u_{i}\right\|
^{2}-\frac{1}{2}\left(  \sum_{j=1}^{N}\operatorname*{tr}K_{ij}\right)
\left\|  u_{i}\right\|  ^{2}+u_{i}^{T}\sum_{j\in\mathcal{I}_{n}}\bar{K}%
_{ij}\Phi_{j}(-\tau_{ij})u_{j}\right)  \label{qq31}%
\end{equation}
provided $\max_{i\in\mathcal{I}_{n}}\left\|  u_{i}\right\|  \leq R$. Let $S$
denote the last summation in (\ref{qq31}), i.e.,%
\begin{align*}
S &  =\sum_{i\in\mathcal{I}_{n}}\sum_{j\in\mathcal{I}_{n}}u_{i}^{T}\bar
{K}_{ij}\Phi_{j}(-\tau_{ij})u_{j}\\
&  =\sum_{i\in\mathcal{I}_{n}}\sum_{j\in\mathcal{I}_{n}}\frac{1}{2}u_{i}%
^{T}[\bar{K}_{ij}\Phi_{j}(-\tau_{ij})+\Phi_{i}^{T}(-\tau_{ji})\bar{K}_{ji}%
^{T}]u_{j}\\
&  =\sum_{i\in\mathcal{I}_{n}}\sum_{j\in\mathcal{I}_{n}}\frac{1}{2}u_{i}%
^{T}[\bar{K}_{ij}\Phi_{j}(-\tau_{ij})+\Phi_{j}^{T}(-\tau_{ij})\bar{K}_{ij}%
^{T}]u_{j}%
\end{align*}
where the last line follows since $\bar{K}_{ij}=\bar{K}_{ji}$ and $\tau
_{ij}=\tau_{ji}$ for all $i,j$, and $\Phi_{i}\equiv\Phi_{j}$ for
$i,j\in\mathcal{I}_{n}$ by our assumptions. Hence from (\ref{Kbar-sym2}),%
\begin{align}
S &  =\sum_{i\in\mathcal{I}_{n}}\sum_{j\in\mathcal{I}_{n}}u_{i}^{T}[\bar
{K}_{ij}\Phi_{j}(-\tau_{ij})]_{s}u_{j}\nonumber\\
&  =\sum_{i\in\mathcal{I}_{n}}\sum_{j\in\mathcal{I}_{n}}\frac{1}{2}\alpha
_{ij}u_{i}^{T}u_{j}\label{qq33}%
\end{align}
where
\begin{equation}
\alpha_{ij}=(\operatorname*{tr}K_{ij})\cos\omega_{n}\tau_{ij}%
-\operatorname*{tr}(JK_{ij})\sin\omega_{n}\tau_{ij}.
\end{equation}
From the identity%
\[
0\leq\left\|  u_{i}\pm u_{j}\right\|  ^{2}=\left\|  u_{i}\right\|
^{2}+\left\|  u_{j}\right\|  ^{2}\pm2u_{i}^{T}u_{j}%
\]
follows the estimate
\[
\left|  u_{i}^{T}u_{j}\right|  \leq\frac{\left\|  u_{i}\right\|  ^{2}+\left\|
u_{j}\right\|  ^{2}}{2}.
\]
Using in (\ref{qq33}) and noting that $\alpha_{ij}=\alpha_{ji}$ gives%
\[
S\leq\frac{1}{4}\sum_{i\in\mathcal{I}_{n}}\sum_{j\in\mathcal{I}_{n}}%
|\alpha_{ij}|\left(  \left\|  u_{i}\right\|  ^{2}+\left\|  u_{j}\right\|
^{2}\right)  =\frac{1}{2}\sum_{i\in\mathcal{I}_{n}}\sum_{j\in\mathcal{I}_{n}%
}|\alpha_{ij}|\left\|  u_{i}\right\|  ^{2}.
\]
Substitution into (\ref{qq31}) yields%
\begin{align*}
\frac{d}{dt}\sum_{i\in\mathcal{I}_{n}}\left\|  u_{i}\right\|  ^{2} &
\leq2\varepsilon\sum_{i\in\mathcal{I}_{n}}\left(  q_{i}-\frac{1}{2}\sum
_{j=1}^{N}\operatorname*{tr}K_{ij}+\frac{1}{2}\sum_{j\in\mathcal{I}_{n}%
}|\alpha_{ij}|\right)  \left\|  u_{i}\right\|  ^{2}\\
&  \leq\varepsilon\lambda\sum_{i\in\mathcal{I}_{n}}\left\|  u_{i}\right\|
^{2}%
\end{align*}
where we have defined
\[
\lambda=\max_{i\in\mathcal{I}_{n}}\left(  2q_{i}-\sum_{j=1}^{N}%
\operatorname*{tr}K_{ij}+\sum_{j\in\mathcal{I}_{n}}|\alpha_{ij}|\right)  .
\]
Thus
\[
\left\|  v_{n}(t)\right\|  ^{2}\leq\left\|  v_{n}(0)\right\|  ^{2}%
\exp(\varepsilon\lambda t)\text{.}%
\]
If condition (\ref{stab-net}) holds, then $\lambda$ is negative, and the first
statement of the theorem is proved.

To prove the second statement, consider the linearization of (\ref{net-avg})
about zero:%
\[
\dot{u}_{i}=\varepsilon\left(  D\bar{f}_{i}(0)u_{i}-\sum_{j=1}^{N}\bar{K}%
_{ij}u_{i}+\sum_{j\in\mathcal{I}_{n}}\bar{K}_{ij}\Phi_{j}(-\tau_{ij}%
)u_{j}\right)  ,\quad i\in\mathcal{I}_{n}.
\]
Recalling that $K_{ii}=0$, and using (\ref{qq25}) and (\ref{Kbar-ij}), the
divergence of the vector field defined by the right hand side is calculated
as
\[
\sum_{i\in\mathcal{I}_{n}}\operatorname*{tr}\left(  D\bar{f}_{i}(0)-\sum
_{j=1}^{N}\bar{K}_{ij}\right)  =\sum_{i\in\mathcal{I}_{n}}\left(
\operatorname*{tr}Df_{i}(0)-\sum_{j=1}^{N}\operatorname*{tr}K_{ij}\right)
\]
which is positive if (\ref{instab-net}) holds. In this case, any sufficiently
small neighborhood of the origin expands in volume under the evolution of the
differential equations (\ref{net-avg}), which proves the instability of the
zero solution.
\end{proof}

In view of the above theorem it is worthwhile to examine the death of a given
network as additional units are connected to it. Suppose that the network
consists of $N$ oscillators with identical (or $\mathcal{O}(\varepsilon)$
close) frequencies, and consider the condition (\ref{stab-net}) for amplitude
death. When an additional oscillator with a different frequency is connected,
the first sum in (\ref{stab-net}) changes by $\operatorname*{tr}(K_{i,N+1})$
while the other terms remain the same. Thus if $\operatorname*{tr}(K_{i,N+1})$
is sufficiently large for $i=1,\dots.N$, then the original set of $N$
oscillators will experience amplitude death after the new connection.
Conversely, if $\operatorname*{tr}(K_{i,N+1})$ is negative for some $i$, then
it is possible for the new connection to ``awaken'' the network from a
previous death state, by satisfying the instability condition
(\ref{instab-net}).

When the newly added oscillator has the same frequency as the rest, the
coupling delay becomes important. For simplicity let us assume that all
oscillators and the coupling conditions between them are identical, so the
indices can be dropped. The death condition (\ref{stab-net}) then has the
form
\begin{equation}
\operatorname*{tr}K-\left|  (\operatorname*{tr}K)\cos\omega\tau
-\operatorname*{tr}(JK)\sin\omega\tau\right|  >2q/N. \label{large}%
\end{equation}
Assuming $q>0$, this condition cannot be satisfied for any $N$ if either
$\sin\omega\tau=0$ or $\operatorname*{tr}K\leq0$. However, if $\sin\omega
\tau\neq0$ and $K$ is symmetric with positive trace, the network will undergo
amplitude death for all sufficiently large network sizes $N$. For nonsymmetric
matrices the same conclusion holds whenever the left hand side of
(\ref{large}) is positive, i.e., if $\operatorname*{tr}K$ is sufficiently
larger than $|\operatorname*{tr}(JK)|$. However, for highly nonsymmetric
matrices the presence of the connection delay may make amplitude death
impossible, as the left hand side of (\ref{large}) becomes negative.

Although physically meaningful in most models, the assumption that there are
no self-connections in the network (\ref{network}) is not an essential
requirement for Theorem \ref{thm:net}. In fact, an investigation of the first
part of the proof shows that (\ref{stab-net}) remains a sufficient condition
for amplitude death even when the matrices $K_{ii}$ are not zero. This
observation points to an interesting connection between the present paper and
previous works on feedback control of oscillations. For if one considers the
system%
\begin{equation}
\dot{x}(t)-Ax(t)-\varepsilon f(x(t))=\varepsilon K(x(t-\tau)-x(t))
\label{qq71}%
\end{equation}
as a special case of the network (\ref{network}) consisting of a single
oscillator, Theorem \ref{thm:net} gives the sufficient condition%
\[
\operatorname*{tr}K-\left|  (\operatorname*{tr}K)\cos\omega\tau
-\operatorname*{tr}(JK)\sin\omega\tau\right|  >2q
\]
for its stability. Since the latter inequality can be satisfied if
$\operatorname*{tr}K$ is large and $\sin\omega\tau\neq0$, one concludes that
for almost all positive values of the delay $\tau$ and sufficiently small
$\varepsilon>0$, the oscillator given by the left hand side of (\ref{qq71})
can be stabilized by a linear delayed feedback of the form $K(x(t-\tau
)-x(t))$, regardless of a precise knowledge about the nonlinearity $f$. This
is in agreement with related results which were obtained for similar systems
in the context of delayed-feedback control of oscillatory behavior
\cite{Atay-JSV98,Atay-AML99,Atay-IJC02,Atay-LNCIS02}.

The possibility that the network may be divided into clusters which separately
experience amplitude death entails the formation of spatial patterns
consisting of oscillating and suppressed units. Such pattern formation is best
studied under more specific assumptions on the network structure; therefore,
it is not treated further in this paper.

\section{Conclusion}

We have presented a mathematical analysis of the phenomenon of amplitude death
for a general class of weakly nonlinear systems which are diffusively coupled
with time delays. Sufficient conditions are derived for the coupled network to
exhibit amplitude death. It is shown that when the individual oscillators are
sufficiently different, it is possible for only parts of the network to be
suppressed, while the rest continues to oscillate. The size of the network may
enhance or inhibit amplitude death depending on the nature of the coupling
matrices, where both the magnitude and the sign of the trace of the matrix
plays an important role. The quantitative conditions presented are rigorously
justified for weakly nonlinear oscillators. In particular, they are
significant when considering networks where each individual unit is near a
Hopf bifurcation. On the other hand, numerical simulations suggest that in
many cases the qualitative aspects of the results remain valid for stronger
nonlinearities as well. In view of the few assumptions required on the network
structure and the individual units, the generality of the results imply that
the death of the network, either in part or as a whole, may be among the
common and robust dynamical behaviors in many practical systems. This should
be significant in studies of interacting periodic processes, such as physical,
biological, and financial cycles.

%
%
\end{document}